\documentclass[aps,prd,groupedaddress]{revtex4}
\def\ga{\mathrel{\raise.3ex\hbox{$>$\kern-.75em\lower1ex\hbox{$\sim$}}}}
\def\la{\mathrel{\raise.3ex\hbox{$<$\kern-.75em\lower1ex\hbox{$\sim$}}}}

\usepackage{amsmath}
\usepackage{amssymb}
\usepackage{bbm}
\usepackage{epsfig}
\begin{document}
\begin{flushright}
SHEP-08-34\\
UT-HET-019
\end{flushright}
\title{Distinctive Higgs Signals of a Type II 2HDM at the LHC}

\author{Shinya Kanemura$^{1}$~\footnote{kanemu@sci.u-toyama.ac.jp}, Stefano Moretti$^{2}$~\footnote{stefano@phys.soton.ac.uk},
Yuki Mukai$^{1}$~\footnote{ mukai@jodo.sci.u-toyama.ac.jp}, Rui
Santos$^{2}$~\footnote{rsantos@cii.fc.ul.pt} and Kei
Yagyu$^{1}$~\footnote{keiyagyu@jodo.sci.u-toyama.ac.jp}}
\affiliation{ $^{1}$ Department of Physics, University of Toyama,
3190 Gofuku, Toyama 930-8555, Japan.\\
$^{2}$  School of Physics and Astronomy, University of Southampton,
Highfield, Southampton SO17 1BJ, UK.}

\date{\today}

\begin{abstract}
\noindent
We perform a numerical analysis of Higgs-to-Higgs decays within a
Type II 2-Higgs Doublet Model (2HDM), highlighting several
channels that cannot occur in its Supersymmetric version, thereby
allowing one to possibly distinguish between these two scenarios.
Our results are compliant with all available experimental bounds
from both direct and indirect Higgs searches and with theoretical
constraints from vacuum stability and perturbative unitarity.
\end{abstract}
%
\maketitle

\section{Introduction}
\label{sec:intr} \noindent
Whilst Supersymmetry (SUSY) \cite{DreesGodboleRoy} provides an
attractive theoretical scenario for physics beyond the Standard
Model (SM) (in its ability to remedy the hierarchy problem,  to
provide a natural dark matter candidate, to enable high scale
gauge coupling unification, etc.), there is to date no evidence
for it. In its miminal (in terms of particle content and gauge
structure) incarnation, the Minimal Supersymmetric Standard Model
(MSSM) \cite{HaberKane}, SUSY is highly predictive though, e.g.,
in the Higgs sector.  Of the initial eight degrees of freedom
pertaining to the two complex Higgs doublets responsible for
Electro-Weak Symmetry Breaking (EWSB) in the MSSM, after the
latter has taken place, three are consumed to give mass to the
$SU(2)\otimes U(1)$ weak gauge bosons, $W^\pm$ and $Z$, so that
five physical Higgs states survive: the CP-neutral ones $h$ and
$H$ ($M_H>M_h$), the CP-odd one $A$ and the charged states
$H^\pm$. In effect, regarding the Higgs sector, the MSSM is
nothing but a 2-Higgs Doublet Model (2HDM) of Type II (in the
nomenclature of Ref.~\cite{catalogue}), whereby SUSY enforces
relations amongst Higgs masses and couplings, on the one hand, and
weak gauge boson masses and interaction parameters, on the other
hand, in such a way that, of the original 7  independent
parameters defining a CP-conserving 2HDM (see below), only two
survive as such in the MSSM \cite{HHG}. These can be taken to be
$\tan\beta$, the ratio of the Vacuum Expectation Values (VEVs) of
the two Higgs doublet fields, and one of the extra Higgs boson
masses. At tree-level, these are the only inputs needed to compute
Higgs masses and couplings to ordinary matter (quarks, leptons and
gauge bosons) as well as Higgs self-couplings. Hence, if SUSY had
chosen all the sparticle masses to be much larger than the SM
objects and the Higgs bosons, so that they are not accessible at
the upcoming Large Hadron Collider (LHC), one intriguing question
to ask would be whether it is possible to distinguish between the
MSSM and a Type II 2HDM on the sole basis of Higgs interactions
with SM matter and/or self-interactions. Or, conversely, whether
it would be possible to dismiss the assumption of such (decoupled)
heavy sparticles, hence of minimal SUSY altogether (aka the MSSM),
from the observation of particular signals in the Higgs sector
alone.

\noindent It is the purpose of this work to prove that this is the
case, exploiting the fact that SUSY prevents some Higgs-to-Higgs
decays, that remain instead possible in a generic Type II 2HDM.
This is primarily connected to the fact that a generic pattern of
Higgs masses, as dictated by SUSY, in the MSSM is the one in which
the $h$ is rather light (in fact, below 130 $GeV$ or so, for heavy
sparticles \cite{Djouadi:2005gj}) whilst the others ($H$, $A$ and
$H^\pm$) are quite heavy and degenerate in mass, the more so the
larger $\tan\beta$~\footnote{To the extent that, for masses of the
heavy Higgs states above 200 $GeV$ or so, a sort of decoupling
occurs between these and the lightest Higgs boson of the MSSM, so
that the latter resembles its SM counterpart.}. In fact, even in
the presence of off-shellness effects \cite{Moretti:1994ds,abdel}
in all decay products in Higgs decay chains in the MSSM,
essentially only the $H\to hh$ and $A\to Zh$ channels are possible
in this scenario~\footnote{While the $H^\pm\to W^\pm h$ channel is
kinematically possible in the MSSM, this only occurs with sizable
rates in a $\tan\beta$ region which is already excluded by
experimental data.}. In contrast, in a Type II 2HDM without SUSY,
many other decays are possible, e.g.:

\vspace*{+0.20cm}
\centerline{$H\to AA,\qquad
H\to H^+H^-, \qquad
H\to  W^\pm H^\mp, \qquad
H\to  ZA,$}

\vspace*{+0.15cm}
\centerline{$A\to  W^\pm H^\mp, \qquad
A\to  ZH,$}
\vspace*{-0.75cm}
\begin{equation}
H^\pm\to W^\pm h, \qquad H^\pm\to W^\pm H \qquad H^\pm\to W^\pm A.
\end{equation}
\noindent While the existence of such different decay patterns in
the two models has been known for some time \cite{HHG}, our
ultimate intention here is to prove that the 2HDM Type II is still
phenomenologically viable in the light of all available
experimental constraints. In practise then, whilst, of course,
these decays could not all occur at the same time, depending on
the Type II 2HDM parameters, a subset of them would be possible,
therefore providing a means of distinguishing between the two
scenarios discussed, further considering that -- under the above
assumption of a heavy SUSY spectrum -- the dominant production
channels of both neutral and charged Higgs bosons in both
scenarios are the same and proceed via interactions with SM
particles \cite{WJSZK,Moretti:2001pp}.

\noindent The plan of the paper is as follows. In the next section
we describe the Type II 2HDM that we will be using and we fix our
conventions. The following section discusses the experimental and
theoretical bounds. We then illustrate our results quantitatively
in section IV. The final section outlines the conclusions.
\section{The two-Higgs doublet model}
\label{sec:eff}
\noindent
We start with a brief review of the 2HDM used in this work. The
potential chosen is the most general, renormalisable and invariant
under $SU(2) \otimes U(1)$ that one can build with two complex
Higgs doublets with a softly broken $Z_2$ symmetry. It can be
written as
\begin{eqnarray}
 V_{\rm 2HDM}&=& \mu_1^2 |\Phi_1|^2+\mu_2^2 |\Phi_2|^2-(\mu_3^2
  \Phi_1^\dagger \Phi_2 + {\rm h.c.})\nonumber\\
&&  + \frac{1}{2} \lambda_1 |\Phi_1|^4
  + \frac{1}{2} \lambda_2 |\Phi_2|^4
 + \lambda_3 |\Phi_1|^2|\Phi_2|^2
  + \lambda_4 |\Phi_1^\dagger \Phi_2|^2
  + \frac{\lambda_5}{2} \left\{(\Phi_1^\dagger \Phi_2)^2 + {\rm h.c.}
                        \right\},
\end{eqnarray}
where $\Phi_1$ and $\Phi_2$ are the two Higgs doublets with
hypercharge $+1/2$ and $\mu_3^2$ is the $Z_2$ soft breaking term.
For simplicity we can take $\mu_3^2$ and $\lambda_5$ to be
real. The doublet fields are parameterised as
 \begin{eqnarray}
  \Phi_i = \left[\begin{array}{c}
            \omega_i^+ \\ \frac{1}{\sqrt{2}}(v_i+h_i + i z_i)
            \end{array}
   \right] \hspace{4mm} (i=1,2),
  \end{eqnarray}
where the VEVs $v_1$ and $v_2$ satisfy $v_1^2+v_2^2 = v^2 \simeq
(246 \hspace{2mm} {\rm GeV})^2$. Assuming CP-conservation, this
potential has 8 independent parameters. However, because  $v$ is
fixed by the $W^\pm$ boson mass, only 7 independent parameters
remain to be chosen, which we take to be $M_{h}$, $M_{H}$,
$M_{A}$, $M_{H^\pm}$, $\tan\beta$, $\alpha$ (the mixing angle
between the two CP-even neutral Higgs states) and $M^2=
\mu_3^2/(\sin \beta \cos \beta)$ (which is a measure of how the
discrete symmetry is broken). The definition of $\alpha$ and
$\beta$ and the relation among physical scalar masses and coupling
constants are shown in Ref.~\cite{KOSY} for definiteness.

%
\noindent In a general 2HDM, the Yukawa Lagrangian can be built in
four different and independent ways so that it is free from
Flavour Changing Neutral Currents (FCNCs). We define as Type II
the model where $\phi_2$ couples to
 up-type quarks and $\phi_1$ couples to down-type quarks and
leptons~\footnote{For completeness, though not dealt with here,
we identify as Type I the model where only the doublet
$\phi_2$ couples to all fermions; in a Type III~\cite{catalogue}
(also known as Type Y) model $\phi_2$
couples to all quarks and $\phi_1$ couples to all leptons; a Type
IV~\cite{catalogue} (also known as Type X~\cite{akty}) model is instead built
such that $\phi_2$ couples to up-type quarks and to leptons and
$\phi_1$ couples to down-type quarks.}. (We
present the Yukawa couplings for a Type II 2HDM in the Appendix.)
\section{Experimental and theoretical bounds}
%
\label{sec:bounds}
%
%
%
\noindent
The results from all
LEP collaborations on topological searches with two Higgs bosons or one
Higgs and one gauge boson  were presented in
Ref.~\cite{Schael:2006cr}. We will use the experimental
limits on the cross sections for $e^+ e^- \rightarrow H_1 Z$ and $e^+
e^- \rightarrow H_1 H_2$, where $H_1$ can be any CP-even Higgs
boson and $H_2$ can be either a CP-even or a CP-odd Higgs boson.
As we are concerned here with a Type II 2HDM, there is a bound
particularly relevant to our analysis which is the one obtained
from the relation involving the following cross sections ($\sigma$) and Branching Ratios (BRs):
\begin{equation}
\frac{\sigma(e^+e^-\to {H_1^{\rm 2HDM} Z)}}{\sigma(e^+e^-\to
{H_1^{\rm SM} Z})} \, {\rm BR} (H_1^{\rm 2HDM} \rightarrow b
\bar{b}) \, = \, \sin^2 (\alpha - \beta) \, {\rm BR} (H_1^{\rm
2HDM} \rightarrow b \bar{b}) \, \, .
\end{equation}
In a Type II 2HDM $h \rightarrow b \bar{b}$ is the dominant decay
for most of the parameter space for $M_h$ below the SM limit. The
two subleading competing decays are $h \rightarrow c \bar{c}$ and $h
\rightarrow \tau^+ \tau^-$. In a Type II 2HDM, the ratio $\Gamma (h
\rightarrow b \bar{b})/\Gamma (h \rightarrow \tau^+ \tau^-)$ is
the SM one. In contrast, one has
\begin{equation}
\frac{\Gamma (h \rightarrow b \bar{b})}{\Gamma (h \rightarrow c
\bar{c})} \, = \, \frac{m_b^2}{m_c^2} \, \tan^2 \alpha \, \tan^2
\beta
\end{equation}
in the limit $m_h >> m_q$. If we then take the case $\alpha
\approx \beta$ and because we always consider $\tan \beta
> 1$ the decay $h \rightarrow b \bar{b}$ becomes even more
dominant in a Type II 2HDM, with respect to the SM case.
Conversely, if $\alpha \approx \beta + \pi/2$, then we recover the
SM ratio. Either way, $h\to b\bar b$ is the dominant decay by an
amount which is at least the corresponding SM ratio of BRs with
respect to the other fermionic decays. Because its dependence on
the other parameters of the model is very mild, this essentially
provides a limit in the $\beta - \alpha$ versus $M_h$ plane. In
particular, it is straightforward to check that when $\sin (\beta
- \alpha) \approx 0.1$ there is essentially no bound on the
lightest Higgs boson mass, while for $\sin (\beta - \alpha)
\approx 0.2$ the limit immediately jumps to $M_h > 75.6$ GeV.

\noindent In this study the masses of the charged Higgs boson and
the CP-odd one will always be above 200 $GeV$ and therefore not
constrained at all by the LEP bounds from direct searches. In some
cases though, the mass of the heavy CP-even Higgs boson will be
allowed to be below 200 GeV. In those instances we have confirmed
that the LEP bounds do not apply to the cases presented in this
work. Concerning $H^\pm$ states, apart from a model independent
LEP bound of $M_{H^\pm}\ga M_{W^\pm}$, D0 and CDF have model
dependent limits on the charged Higgs mass (see
Ref.~{\cite{charged}) from top decays, but again these are below
the range of $H^\pm$ masses discussed in this work. No other
experimental bounds exist from direct searches for the set of
parameters that we will present.

\noindent Other than limits from direct searches for Higgs bosons,
there are indirect constraints from precision observables, from
both LEP and SLC. New contributions to the $\rho$ parameter
stemming from Higgs states \cite{Rhoparam} have to comply with the
current limits from precision measurements \cite{pdg4}: $
|\delta\rho| \la 10^{-3}$. There are limiting cases though,
related to an underlying custodial symmetry,  where the extra
contributions to $\delta\rho$ vanish. In this study we will
consider two such particular cases: (A) the one where
$M_{H^\pm}^{} = M_A^{}$ and (B) the one where
$M_{H^\pm}^{}=M_H^{}$ with $\sin(\beta-\alpha)=1$. These parameter
choices correspond to the case in which the custodial symmetry
($SU(2)_L \otimes SU(2)_R \to SU(2)_V$) is preserved in the Higgs
potential, so that the latter can be written  in terms of
${\rm Tr}(M_i M_i^\dagger)$ with $M_i=(i \tau_2 \Phi_i^\ast,
\Phi_i)$ where the $M_i$'s ($i=1,2$) are translated as $M_i \to
M_i'=g_L^\dagger M_i g_R^{}$ with $g_{L,R}^{} \in SU(2)_{L,R}$ (A)
or ${\rm Tr}(M_{21} M_{21}^\dagger)$ and ${\rm det}(M_{21})$, with
$M_{21}=(i \tau_2 \Phi_2^\ast,\Phi_1)$, where $M_{21}$ is
translated as $M_{21} \to M_{21}'=g_L^\dagger M_{21} g_R^{}$ with
$g_{L,R}^{} \in SU(2)_{L,R}$ (B), respectively
\cite{pomarol-vega}. In Fig.~\ref{rho}, we show the allowed region
under the $\rho$ parameter constraint in several scenarios which
are relevant to our later discussions. Furthermore, it has
recently been shown in Ref.~\cite{Oslandk} that, for a Type II
2HDM, data on $B\to X_s \gamma$ imposes a lower limit of
$M_{H^\pm} \ga 250$\,GeV, which is essentially $\tan\beta$
independent. Other experimental constraints on a Type II 2HDM come
from the results on $(g-2)_\mu$ (the muon anomalous magnetic
moment) \cite{g-2}, $R_b$ (the $b$-jet fraction in $e^+e^-\to
Z\to$ jets) \cite{LEPEWWG,SLD}, the decay $B^+ \to \tau^+
\nu$~\cite{btaunu}, $B_q \bar{B_q}$ mixing~\cite{Oslandk} and the
$\tau$ leptonic decay \cite{tau-ldecay}. In general, bounds from
these observables can be important for relatively small values of
$M_{H^\pm}^{}$ and large $\tan\beta \ga10$
\cite{g-2,LEPEWWG,SLD,btaunu, tau-ldecay,chankowski,cheung}.
Values of $\tan \beta$ smaller than $\approx 1$ are disallowed
both by the constraints coming from $Z \rightarrow b \bar{b}$ and
from $B_q \bar{B_q}$ mixing.

\begin{figure}
 \includegraphics[height=3.0in,width=3.0in,angle=0]{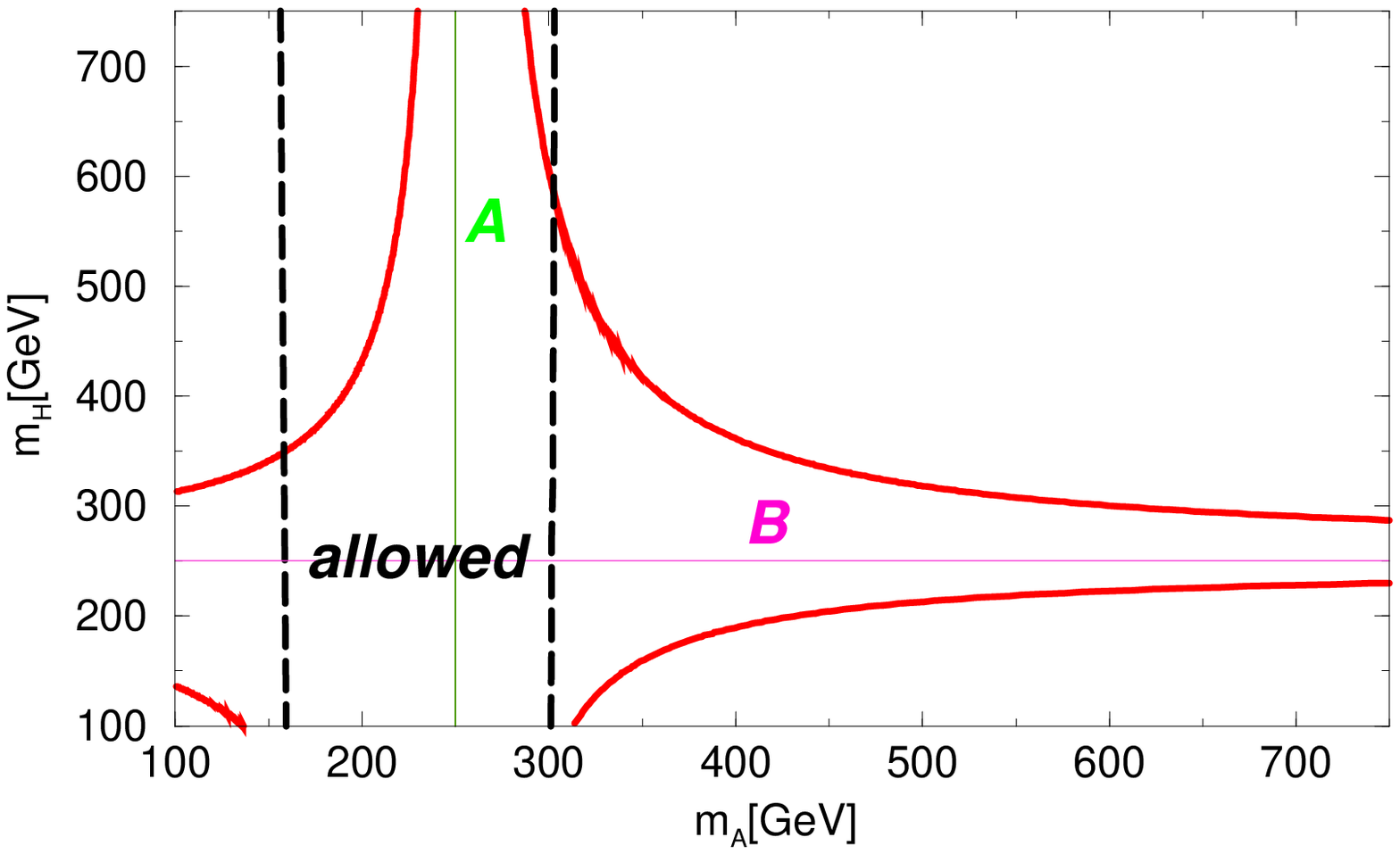}
 \includegraphics[height=3.0in,width=3.0in,angle=0]{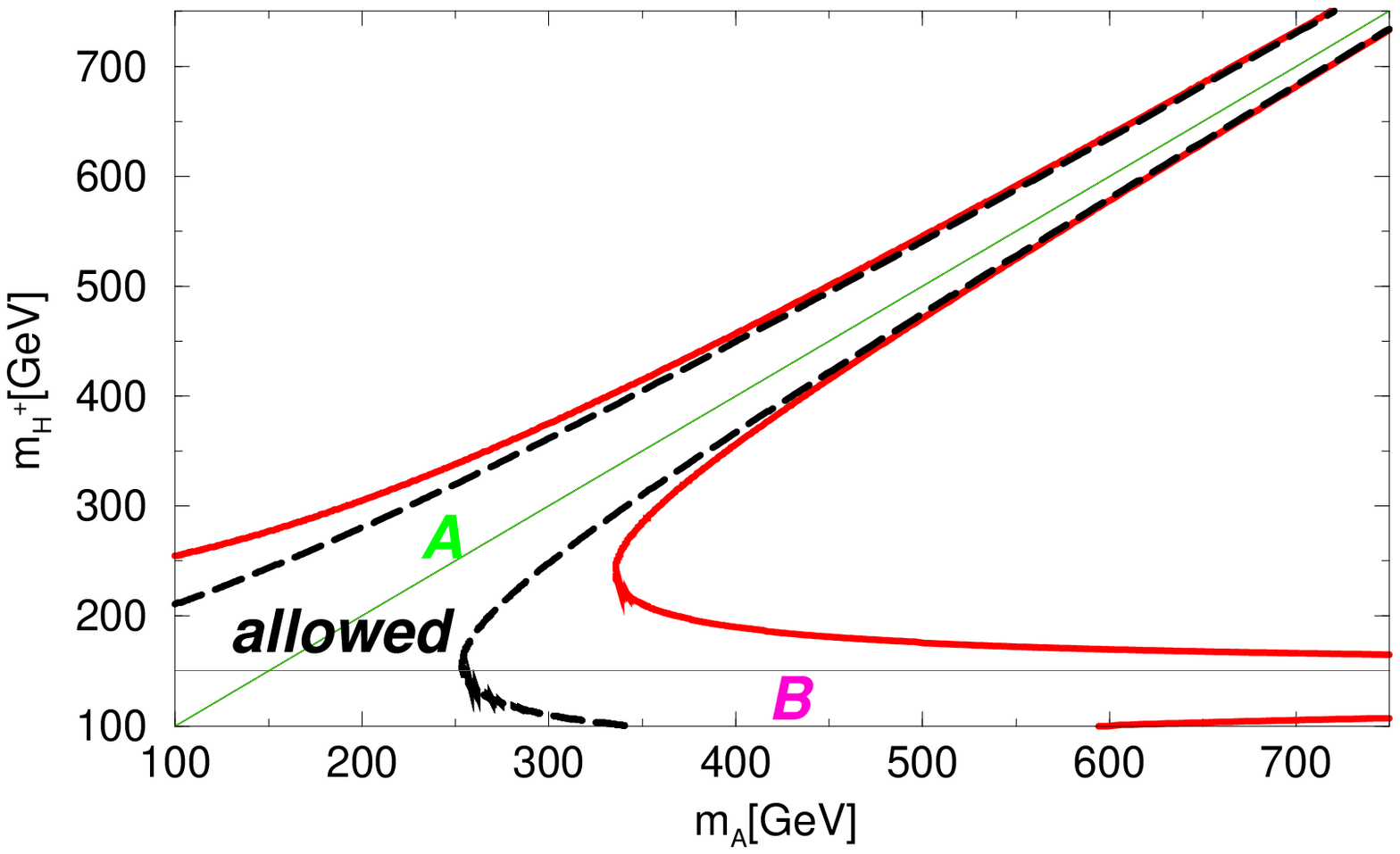}
  \caption{Allowed regions under the $\rho$ parameter data (at $2\sigma$ level). [Left figure]
 The area
 surrounded by solid curves corresponds to the allowed regions for the
 case of $M_h=120~{\rm{GeV}},~M_{H^\pm}^{}=250~{\rm{GeV}},~\sin(\beta-\alpha)=1$. The
 one surrounded by dashed curves corresponds to the allowed regions for
 the case of
 $M_h=50~{\rm{GeV}},~M_{H^\pm}^{}=250~{\rm{GeV}},~\sin(\beta-\alpha)$=0.1.
 [Right figure]  The area surrounded by solid curves corresponds to the
 allowed regions for the case of $M_h=120~{\rm{GeV}},~M_{H}^{}=150~{\rm{GeV}},~\sin(\beta-\alpha)$=0.9. The one
 surrounded by dashed curves corresponds to the allowed regions for the
 case of $M_h=50~{\rm{GeV}},M_{H}^{}=150~{\rm{GeV}},~\sin(\beta-\alpha)$=0.1. }
  \label{rho}
\end{figure}

\noindent Concerning theoretical constraints we will take all
masses $M_h$, $M_H$, $M_A$ and $M_{H^\pm}$ to be below 700 GeV.
This is a consequence of tree-level unitarity
bounds~\cite{unit1,abdesunit} in the limit $M=0$.
Furthermore, the most general set of conditions for the Higgs
potential to be bounded from below are~\cite{vac1}
\begin{equation}
\lambda_1  > 0\;, \quad \quad \lambda_2 > 0\;,
 \quad \quad  \sqrt{\lambda_1\lambda_2 } + \lambda_{3}  + {\rm{min}} \left( 0
, \lambda_{4}-|\lambda_{5}|
 \right) >0.
\label{vac}
\end{equation}
Recently, it was in fact proven that these are necessary and
sufficient conditions to assure vacuum stability of the potential
at tree level~\cite{Ivanov:2006yq}. Vacuum stability against
charge breaking is also built into a Type II 2HDM model, as a
non-charge breaking minimum, when it exists, is always the global
one in any 2HDM~\cite{Ferreira:2004yd}. Finally, according to
\cite{berger}, perturbativity for the top and bottom Yukawa
couplings forces $\tan\beta$ to lie in the range $0.3\leq
\tan\beta \leq 100$, though it turns out that, from enforcing
perturbativity also on the $\lambda_i$'s, moderate values of
$\tan\beta$ ($\tan \beta \sim {\cal O}(1)$) are preferred,
especially for $M \sim 0$ GeV.

\noindent In the following, we will consider $M_{H^\pm}$ to be 250
$GeV$ or larger and take $\tan\beta \sim 1$--$3$, so that all the
Type II 2HDM scenarios presented are free from these bounds. We
will show that even with such small values of $\tan \beta$ a lot
of the parameter space is already excluded. This does not mean
that larger values of $\tan \beta$ are not allowed but rather that
they are less likely to occur. Choosing a large $\tan \beta$
forces a very particular set of values for the remaining free
parameters if one is to comply with all constraints. Therefore it
may seem we are just scanning over small corners of the 2HDM
parameter space. This is not the case. The 2HDM is already tightly
constrained by experiment and it becomes severely constrained when
one adds the theoretical bounds, especially those from
perturbative unitarity. The EW precision data are also very
restrictive. As shown in Fig.~\ref{rho} (left), where we have
fixed the mass of the charged Higgs at 250 $GeV$, a vast range of
values of the heavy CP-even Higgs is possible. The allowed region
for the pseudo-scalar masses depends on the values of $\beta
-\alpha$: when $\sin (\beta - \alpha) \approx 1$ (full curve) all
values of $M_A$ are allowed provided $M_H$ is close to
$M_{H^\pm}=250 \, GeV$. As we move away from $\sin (\beta -
\alpha) = 1$ (dashed curve) the range of allowed values of the
pseudo-scalar mass shrinks to a region around the value of
$M_{H^\pm}=250 \, GeV$ as wide as the precision measurements
permit. On the right plot we see exactly the same trend but now in
the $M_{H^\pm}$ vs $M_A$ plane. In the plots shown in the
following sections we choose the exact limits that cancel the
$\rho$ parameter contribution. We note however that we have
explicitly checked that varying the values of the masses complying
with these constraints does not produce a qualitative change in
our analysis. In most cases not even a quantitative change is
noticed. We end this section by underlining once more that we are
not focusing on a small corner of the 2HDM parameter space.
Considering all constraints a 2HDM Type II is subject to and the
above discussion it is clear that we are spanning the entire
parameter range allowed. When we focus on a definite limit, like
$M_{H^\pm} = M_A$, it is for illustrative purposes only.

\section{Decays}
\noindent Let us start by saying that all the widths and BRs
presented in this work are calculated at tree-level except for the
decays to $gg$ (plus $\gamma\gamma$ and $Z\gamma$, not visible
though in our plots) which are one-loop processes at the lowest
order. However, we take into account the leading one-loop QCD
corrections to Higgs to quark-antiquark decays by computing
one-loop running masses for the (anti)quarks in the Yukawa
couplings, evaluated at the decaying Higgs mass.

\subsection{$A$ decays}
\begin{figure}[h!]
\centering
\includegraphics[height=3.0in]{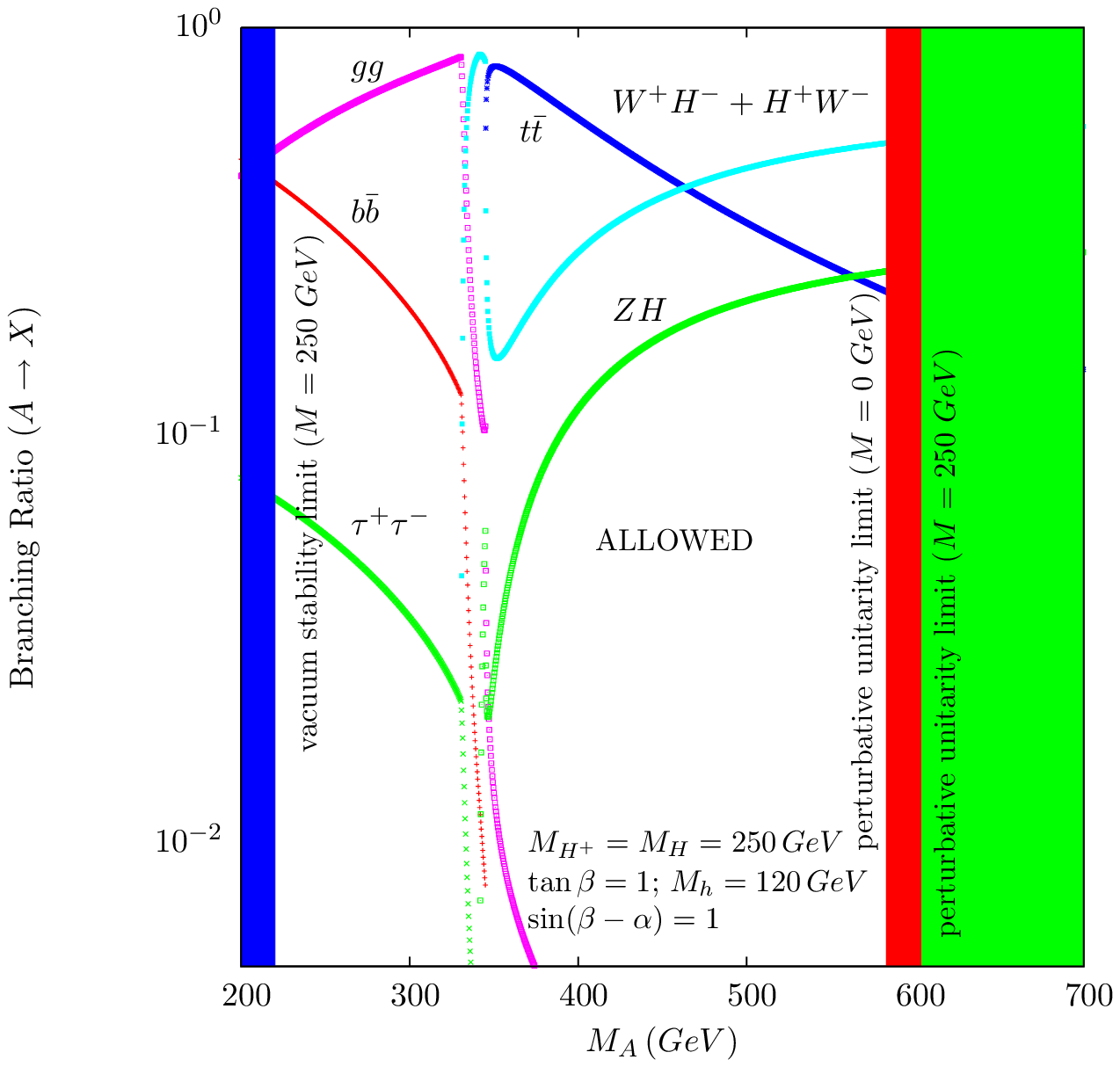}
\includegraphics[height=3.0in]{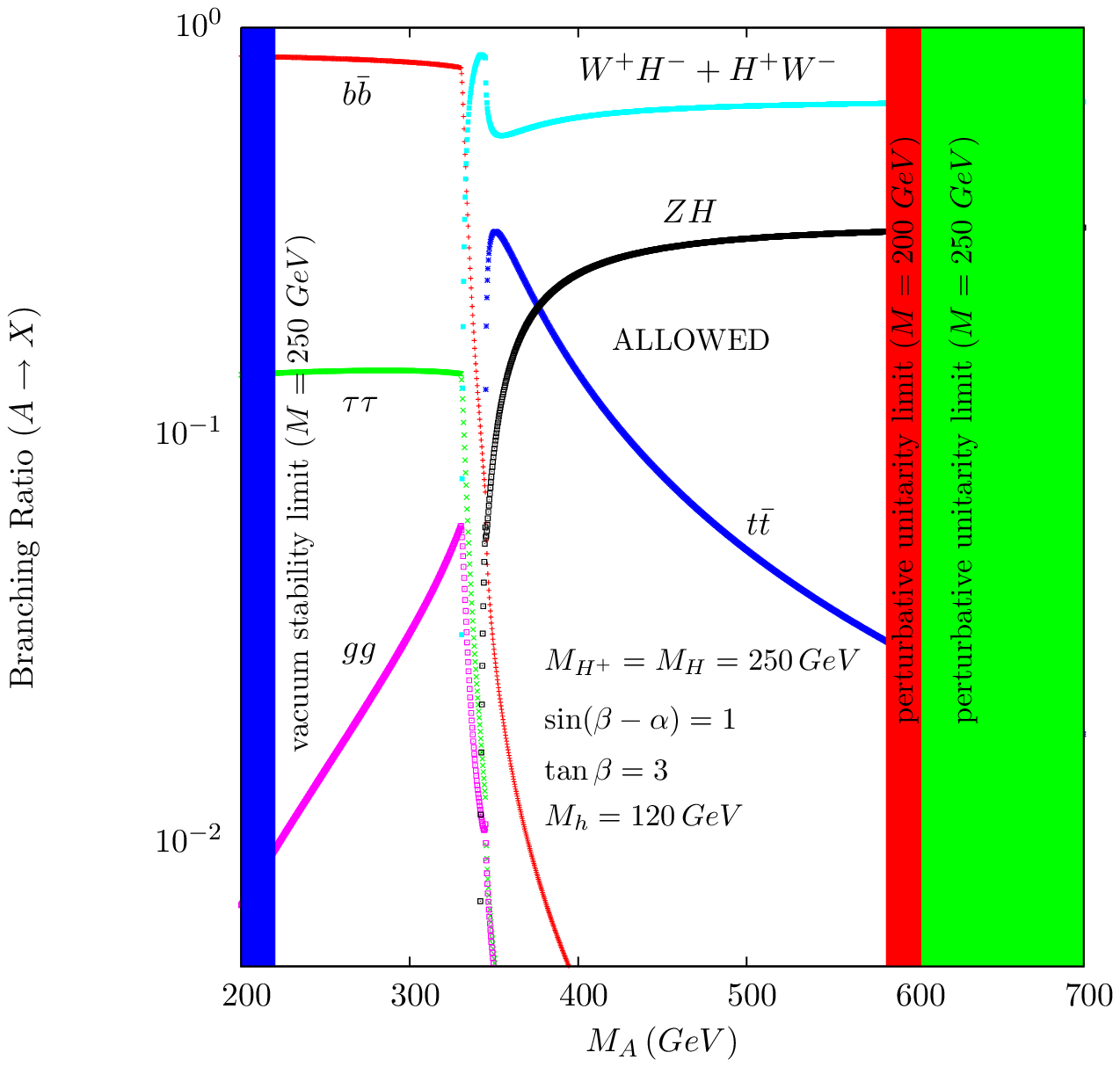}
\caption {$A$ decays for $M_h=120$ GeV, $M_H=M_{H^\pm}=250$ GeV
and $\sin(\beta-\alpha)=1$. On the left is the plot for $\tan
\beta = 1$ while on the right we set $\tan \beta = 3$. We show the
perturbative unitarity limits for $M=0$ $GeV$ and $M=250$ $GeV$ on
the left plot and for $M=200$ $GeV$ and $M=250$ $GeV$ on the right
plot. For $M=0$ GeV and $\tan \beta = 3$ all parameter space is
excluded. We also show the vacuum stability limits for $M=250$
$GeV$. For $M=200$ $GeV$ and below all parameter space is allowed
in what concerns vacuum stability.} \label{plotAdecays}
\end{figure}
\noindent For the CP-odd scalar and in the mass region chosen,
$\sin(\beta-\alpha)$ has to be very close to 1, in order for the
model  to be consistent with experimental data. Even a value of
$\sin(\beta -\alpha)=0.9$ is enough to violate precision
measurements via the $\rho$ parameter. Therefore the decay $A
\rightarrow Zh$ is not allowed. We take $M_h=120$ GeV but the
CP-odd Higgs boson profile does not depend on the light CP-even
Higgs boson mass. We have chosen $M_H=M_{H^\pm}=250$ $GeV$ due to
the $B \rightarrow X_s \gamma$ bound and to the $\rho$ constraint.
Fig.~\ref{plotAdecays} illustrates the decay patterns of the
CP-odd Higgs state for two values of $\tan\beta$. Two comments are
in order here. Firstly, notice that to choose larger values for
$M_H$ and $M_{H^\pm}$ would only have the effect that the
corresponding channels would open later. Secondly, the dependence
on $\tan \beta$ is generally as follows: the larger $\tan \beta$
the more suppressed the decay into $t \bar{t}$ and consequently
the CP-odd Higgs state decays more and more into other Higgs
bosons as soon as the corresponding channels are kinematically
allowed. All decay channels shown in these plots do not depend on
$M$, as no Higgs self-couplings are involved in those processes.
However,
 both the perturbative unitarity and the vacuum
stability bounds depend on the value chosen for $M$. The excluded
regions due to the the above constraints for the three values of
$M=$ 0, 200 and 250 $GeV$ are shown in Fig.~\ref{plotAdecays}.
Note that the smaller the parameter $M$ is the smaller $\tan
\beta$ has to be to avoid the perturbative unitarity limit. On the
contrary, the smaller $M$ is, the less constrained is the
parameter space from the vacuum stability conditions (for $M=0$
GeV no bounds apply).

\noindent In the MSSM, for $\tan \beta = 3$, the pseudo-scalar
decays mainly to fermion pairs, $b\bar{b}$ in the low mass region
and $t \bar{t}$ when the channel becomes kinematically allowed in
most of the studied scenarios~\cite{djouadi08}. There is a
situation where the MSSM and the general 2HDM are hard to
distinguish. The branching ratio of $A \to ZH$ can be very similar
in both models when we compare the so-called intermediate-coupling
regime of the MSSM ($\tan \beta \approx 3$ and $H/A$ masses below
the the $t \bar{t}$ threshold) with a 2HDM with a large charged
Higgs mass so that the the decay to $W^+ H^-$ is forbidden. In
Fig.~\ref{plotAwidth} we show the total $A$ width for the two
situations presented in Fig.~\ref{plotAdecays}. When the $A$
decays mainly to $b\bar{b}$ or $gg$ the width is negligible but,
when  the $t\bar{t}$, the other Higgs or Higgs plus gauge boson
channels open, $\Gamma_A$ grows rapidly reaching 100 $GeV$ for
$M_A=700$ $GeV$. Note that the theoretical constraints shown in
Fig.~\ref{plotAdecays} are not shown again in this plot.

\begin{figure}[h!]
  \begin{center}
    \epsfig{file=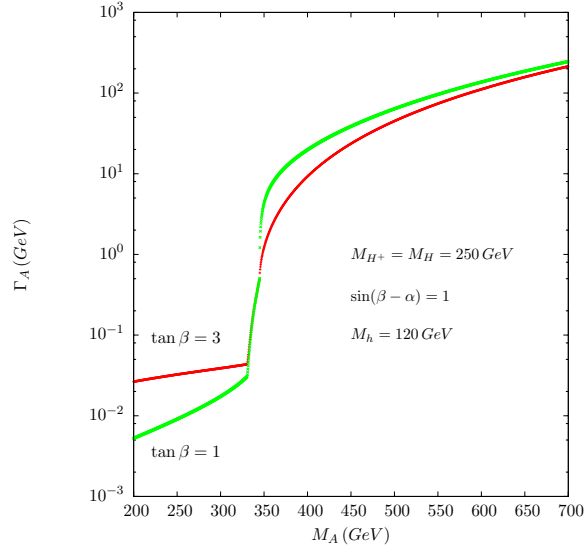,width=8 cm}
    \caption{Total width of the CP-odd Higgs state $A$ for
 for the parameter values presented
    in Fig.~\ref{plotAdecays}.}
    \label{plotAwidth}
  \end{center}
\end{figure}
%
%
%
\subsection{$H$ decays}
%
\begin{figure}[h!]
\centering
\includegraphics[height=3.0in]{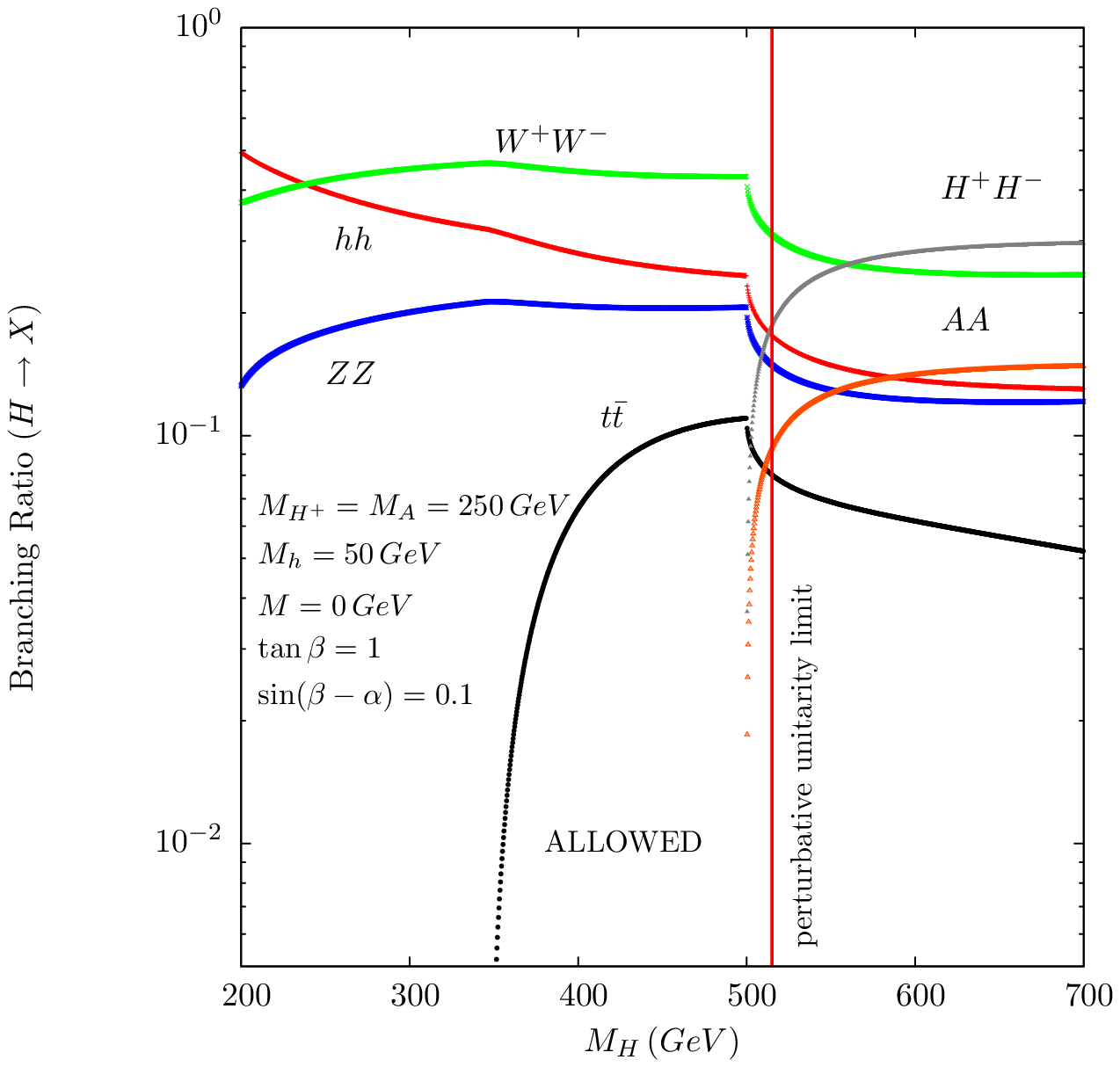}
\includegraphics[height=3.0in]{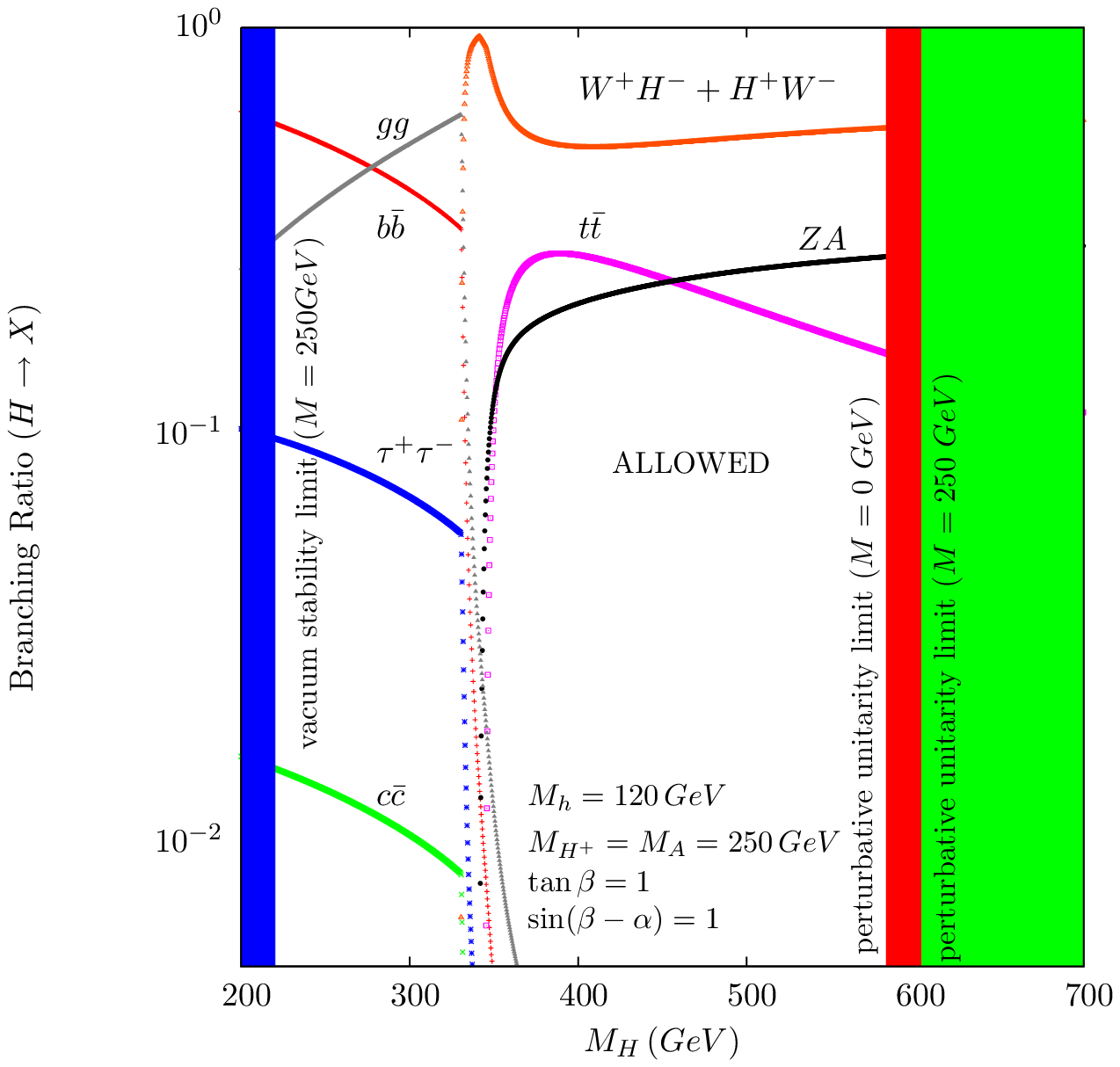}
\caption {$H$ decays for $M_A=M_{H^\pm}=250$ $GeV$ and $\tan \beta
= 1$. On the left is the plot for $M_h=50$ $GeV$ and
$\sin(\beta-\alpha)=0.1$
 while on the right we have $M_h=120$ $GeV$ and
$\sin(\beta-\alpha)=1$. We show the perturbative unitarity and the
vacuum stability limits, for $M=0 \, GeV$ for the left plot and
for $M=0 \, GeV$ and $M=250 \, GeV$ for the right plot.}
\label{plotHdecays1}
\end{figure}

\noindent In this section we deal with the decays of the heavier
CP-even Higgs boson. New contributions to the $\rho$ parameter are
avoided by setting $M_A=M_{H^\pm}=250$ $GeV$. We note once more
that, as shown in Fig.~\ref{rho}, this limit can be relaxed. As a
consequence the dominant channel can be either the one with a
charged Higgs or the one with a pseudo-scalar boson depending on
the relation between the respective masses. Again, the constraint
from $B \rightarrow X_s \gamma$ is used. Then, we distinguish
between two extreme situations. The first one, shown in
Fig.~\ref{plotHdecays1} (left), is for $\sin(\beta-\alpha)=0.1$.
In this case, the $H$ couplings to the gauge bosons are close to
the SM ones. Due to the small value of $\sin(\beta-\alpha)$ the
mass bound on the light Higgs can easily be evaded and we choose
the mass $M_h = 50$ $GeV$ (though smaller masses are also
allowed). This is also the limit where the $H$ couplings to $W^\pm
H^\mp$ and $ZA$ are very small. In this left plot we can
distinguish two interesting regions with new physics signatures.
For small $H$ masses, $H \rightarrow hh, \, W^+ W^-, \, ZZ$ can be
the dominant decays. For large $H$ masses, the decays to a pair of
charged Higgs bosons and to a pair of CP-odd Higgs states dominate
as soon as they are kinematically allowed (though
for a small $M_H$ interval). The larger $\tan \beta$
is the more dominant these decays become. We also present the
perturbative unitarity limit for $M=0 ~ GeV$. We still present
the decays that are above the perturbative unitarity limit for two
reasons. Firstly, because as $M$ grows the allowed $M_H$ region also grows
(although for, for example, for $M=200
\, GeV$, the $H\to hh$ channel is negligible and in the low $H$ mass
region $H \rightarrow W^+ W^-$ always dominate) and the decay to
two charged Higgs and/or two pseudo-scalars are then allowed
over a much larger $M_H$ interval.
Secondly, because, had we chosen a lower value for $M_A$ and
$M_{H^\pm}$, the corresponding decays would have opened for lower
$M_{H}$ values, hence well within the region allowed by perturbative
unitarity.

The other extreme situation is shown on the right hand side of
Fig.~\ref{plotHdecays1} and occurs for $\sin(\beta-\alpha)=1$. In
this limit, the $H$ couplings to the gauge bosons are exactly
zero. Decays to the two light Higgs states are also suppressed but
they could still play a role if the soft breaking parameter is
different from zero. This is the limit where $H$ couplings to
$W^\pm H^\mp$ and $ZA$ are largest. Again we show perturbative
unitarity and vacuum stability constraints for this case. Notice that the
choice of $\tan \beta =1$ is heavily imposed by these bounds:
for $\tan \beta =2$ and for the same set of parameters shown in
the plot, the heavy CP-even Higgs mass is forced to be below
$\approx \, 350\, GeV$. Even if $M$ is raised to $200 \, GeV$ the
bound only grows to $\approx \, 400 \, GeV$.
\begin{figure}[t!]
  \begin{center}
    \epsfig{file=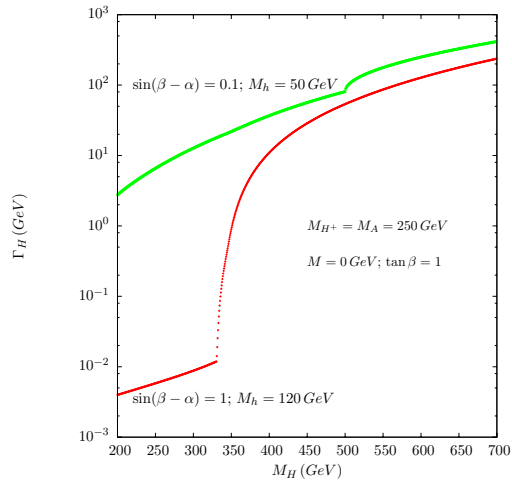,width=7 cm}
    \caption{Total width of the CP-even heavy Higgs state $H$ for the parameter
 values presented
    in Fig.~\ref{plotHdecays1}.}
    \label{plotHwidth}
  \end{center}
\end{figure}
%
%
In Fig.~\ref{plotHwidth} we show the total $H$ width for the two
situations presented in Fig.~\ref{plotHdecays1}. When $M_h = 50
GeV$, left plot in Fig.~\ref{plotHdecays1}, the heavy Higgs is
allowed to decay to other Higgs and gauge bosons and that is why
the $H$ width is SM-like for the same mass. In the other scenario
the heavy CP-even Higgs is not allowed to decay to gauge bosons.
It decays to two gluons and fermion pairs in the low mass region.
Therefore the $H$ width is much smaller. As soon as channels with
Higgs and gauge boson open both widths converge to similar values.

\noindent In the MSSM, for $\tan \beta > 1$, the heavy Higgs
decays mainly to fermion pairs $b \bar{b}$ and then $t \bar{t}$ in
the decoupling regime. Outside this regime there are two
particular cases where distinguishing between both models will be
hard. The first one is the anti-decoupling regime ($\tan \beta
\gtrsim 10$ and $M_A \lesssim M_h^{max}$) where, if kinematically
allowed, the $H \to hh$ can be the dominant decay channel. Hence,
a 2HDM Higgs as presented in the left plot of
Fig.~\ref{plotHdecays1} can be mistaken by such a MSSM heavy
Higgs. The same is true for the above described
intermediate-coupling regime where $Br(H \to hh)$ reaches 60 \%
for a significant heavy Higgs mass region. For a detailed
discussion see~\cite{djouadi08}.

\subsection{$H^\pm$ decays}
\noindent
This section is dedicated to charged Higgs boson decays. Again, we
choose $M_{H^\pm}=M_{A}$ to avoid the constraints from the $\rho$
parameter. Once more we distinguish between two extreme cases
regarding the value of $\sin (\beta -\alpha)$, which is the
parameter that regulates the $H^\pm$ coupling to other Higgses and
gauge bosons. When $\sin (\beta -\alpha)$ is such that the LEP
bounds can be avoided there are mainly two competing decays for
the allowed charged Higgs boson mass region: $H^+ \rightarrow t
\bar{b}$ and $H^+ \rightarrow W^+ h$. In Fig.~\ref{plotHpmdecays1}
we show the charged Higgs BRs for $\sin (\beta-\alpha) =0.1$ and
$M_h=50$ $GeV$ for two values of $\tan \beta$. It is clear that
the decay $H^+ \rightarrow W^+ h$ is always important and becomes
dominant for large values of $\tan \beta$.
\begin{widetext}
\begin{figure}[h!]
\centering
\includegraphics[height=2.9in]{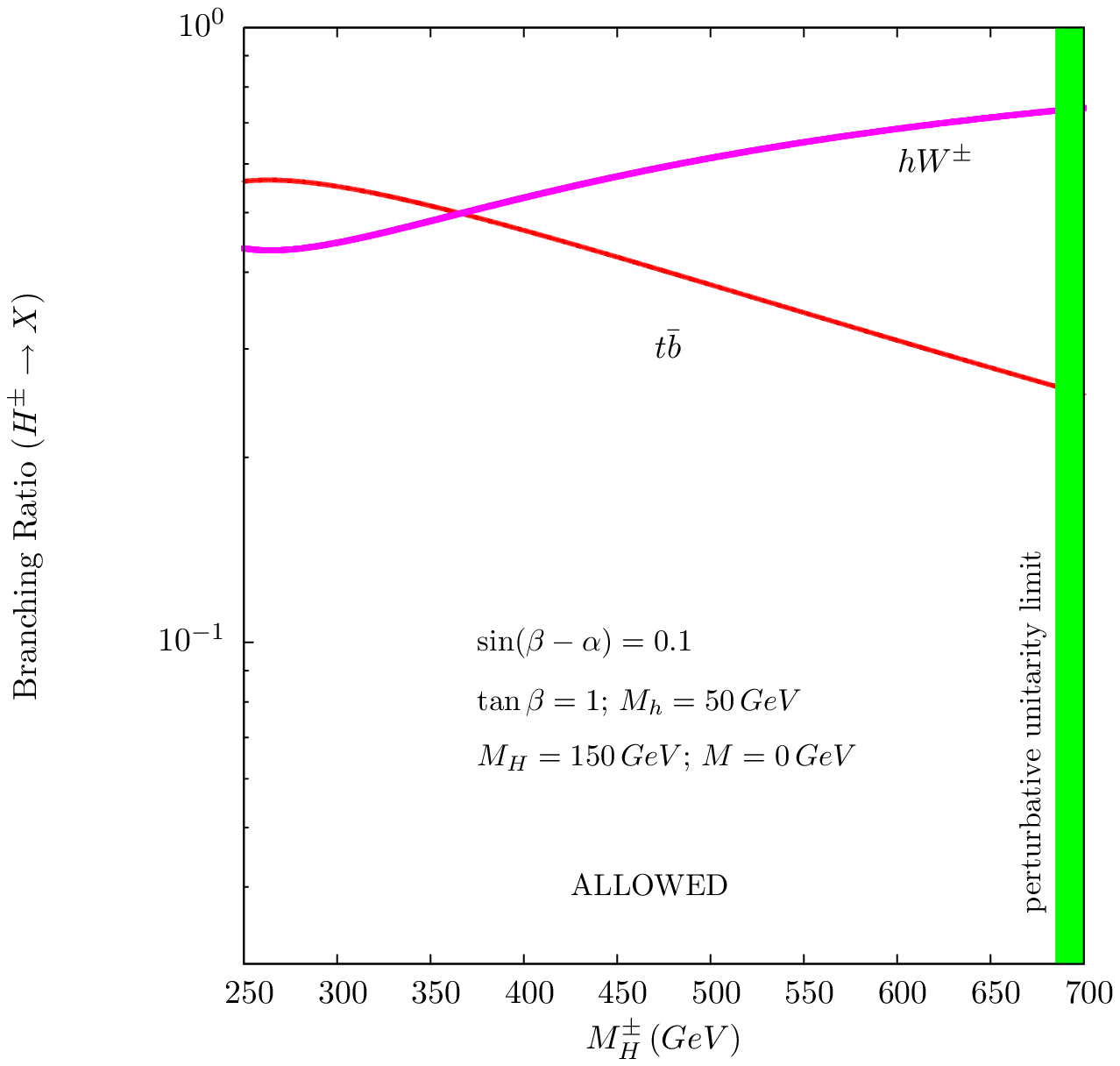}
\includegraphics[height=2.9in]{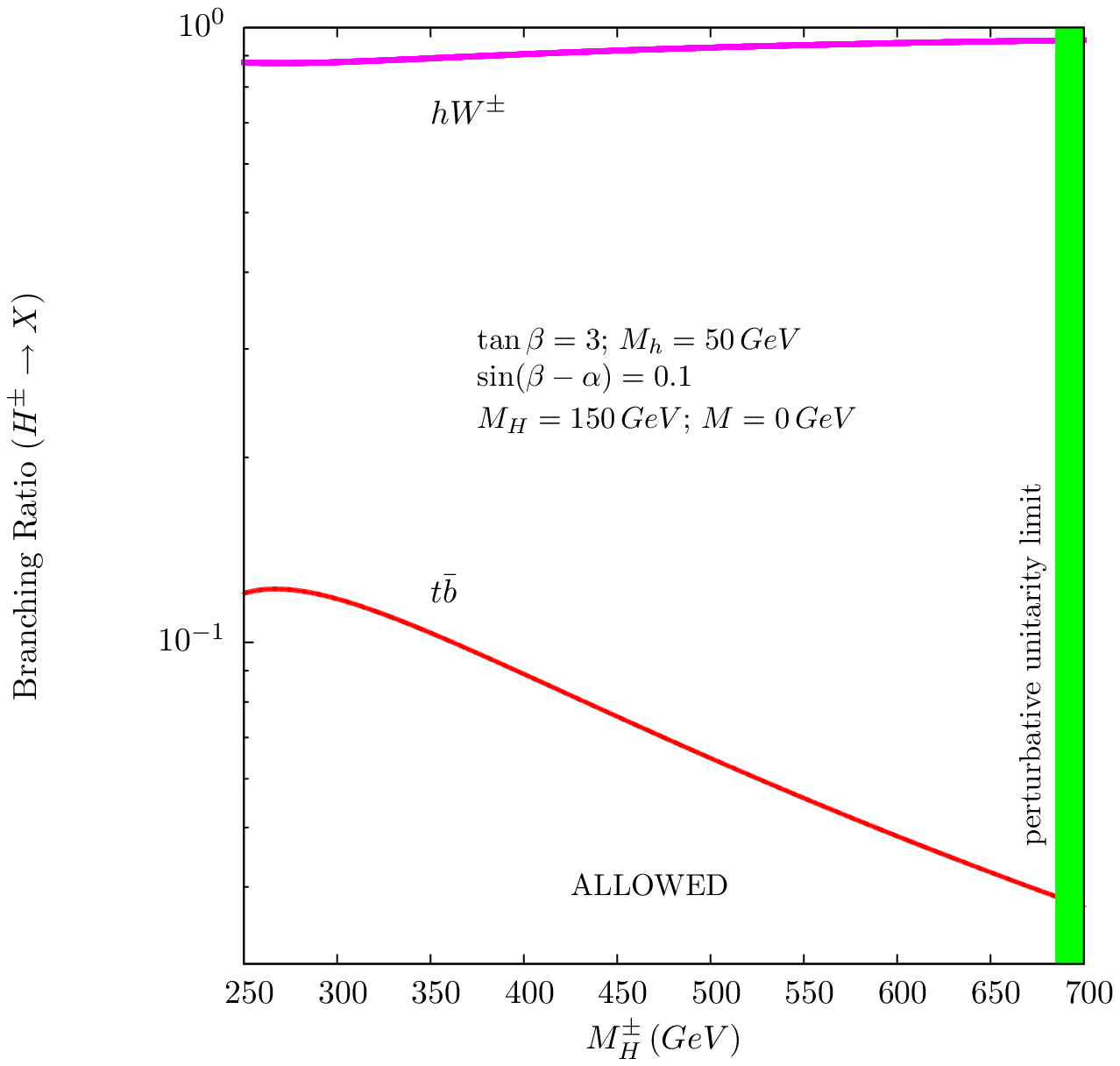}
\caption {$H^\pm$ decays for $M_h=50$ $GeV$, $M_H=150$ $GeV$,
$M_A=M_{H^\pm}$ and $\sin(\beta-\alpha)=0.1$. On the left is the
plot for $\tan \beta = 1$ while on the right we set $\tan \beta =
3$. Perturbative unitarity limits are shown.}
\label{plotHpmdecays1}
\end{figure}
\end{widetext}
The other extreme case, $\sin (\beta-\alpha) =0.9$, is plotted in
Fig.~\ref{plotHpmdecays2}. In this case the $H^\pm$ coupling to
the heavier CP-even Higgs boson becomes dominant relative to the
light Higgs case and the decay $H^+ \rightarrow W^+ H$ is now the
leading one for large values of $\tan \beta.$ The $\rho$
constraint could alternatively be enforced by $M_{H^\pm} \approx
M_{H}$ because $\sin (\beta-\alpha) \approx 1$ as can be seen from
the left plot in Fig.~\ref{rho}. In that scenario the $H W^+$
final state would be replaced by $H^+ \to A W^+$ which is
independent of the value chosen for $\tan \beta$.
\begin{figure}[h!]
\centering
\includegraphics[height=2.9in]{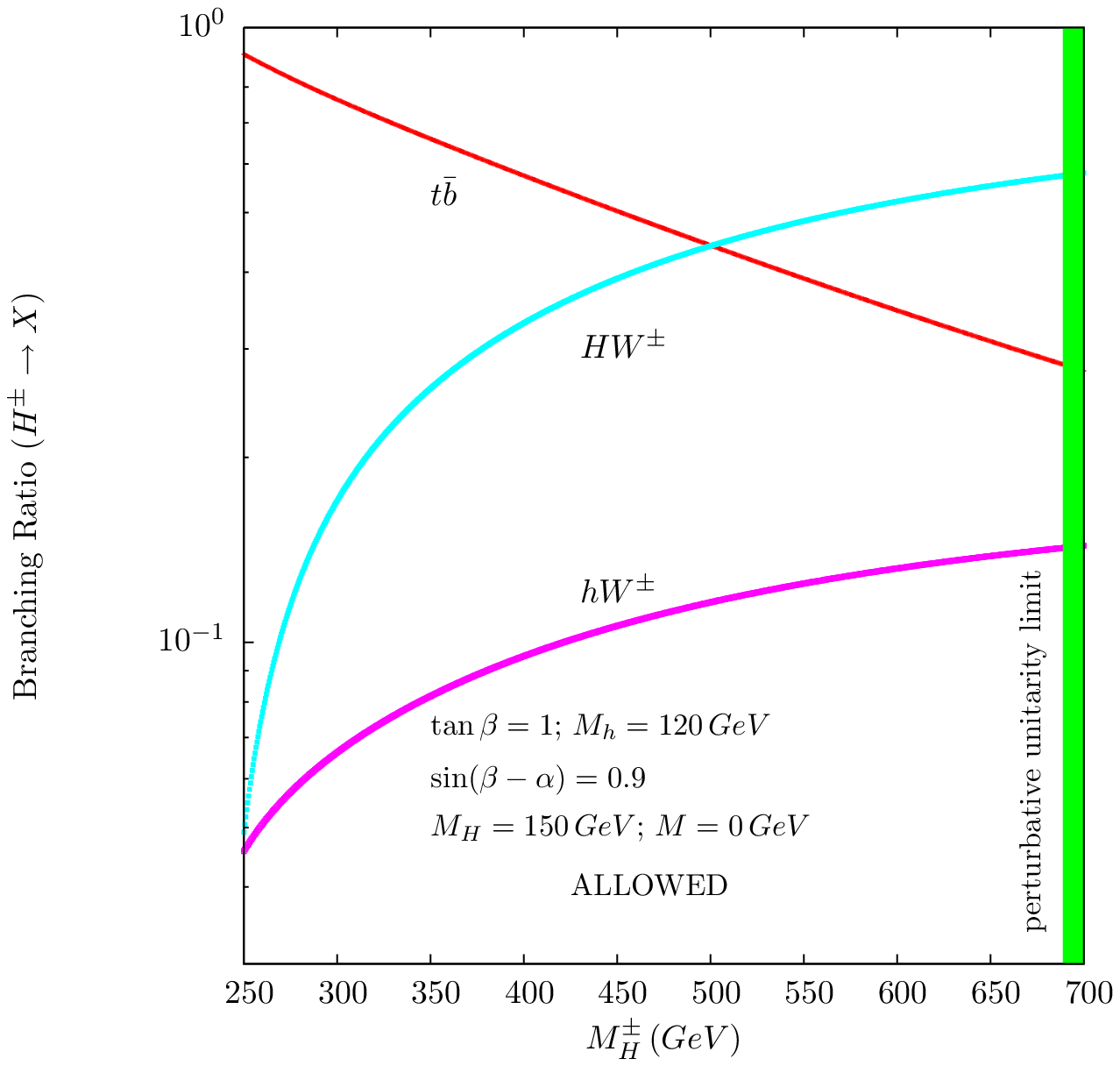}
\includegraphics[height=2.9in]{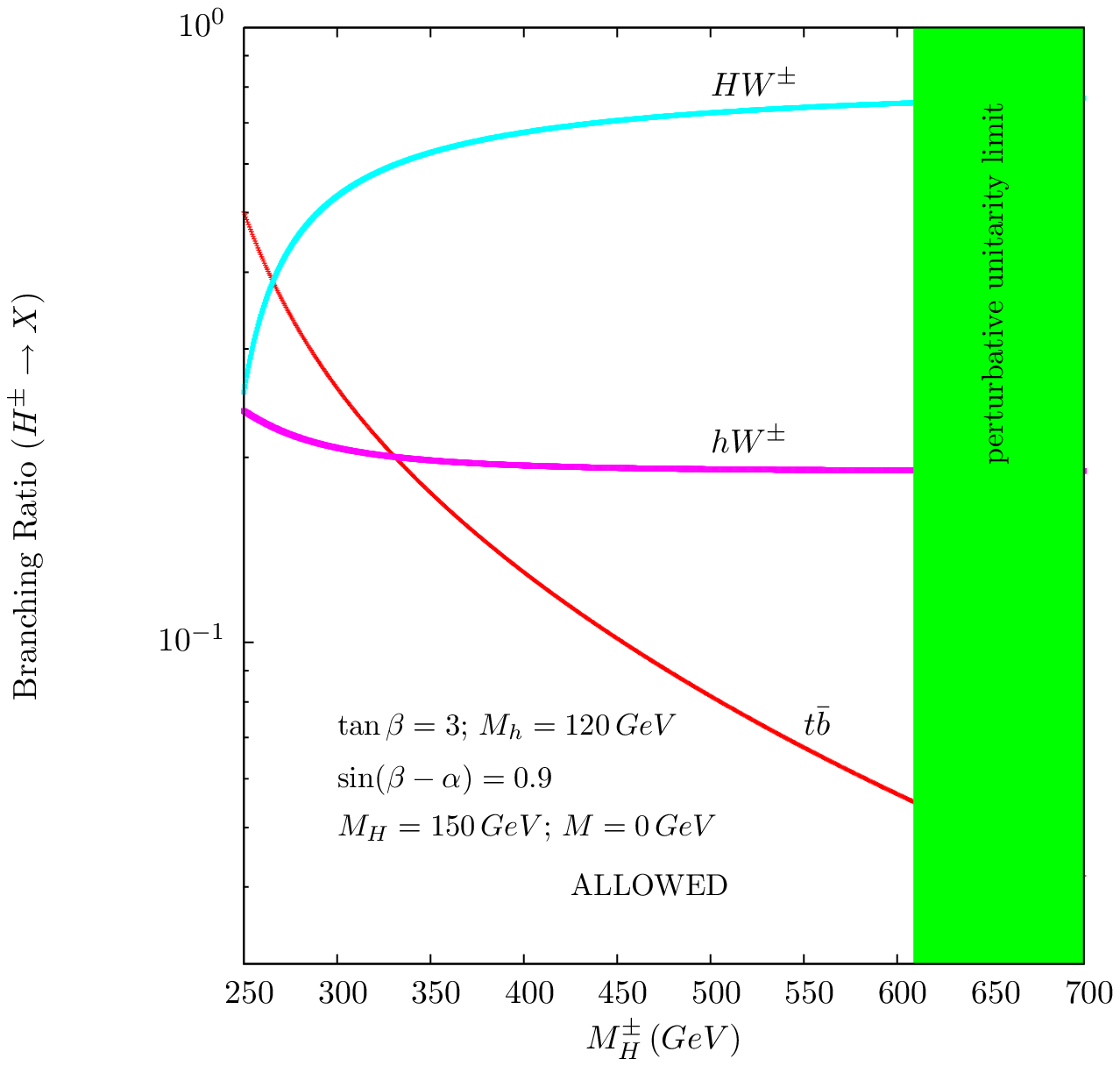}
\caption {$H^\pm$ decays for $M_h=120$ $GeV$, $M_H=150$ $GeV$,
$M_A=M_{H^\pm}$ and $\sin(\beta-\alpha)=0.9$. On the left is the
plot for $\tan \beta = 1$ while on the right we set $\tan \beta =
3$.  Perturbative unitarity limits are shown.}
\label{plotHpmdecays2}
\end{figure}
Again,  we take $M_{H}=150$ $GeV$, though this value has no
bearing on the final result except when we are in a region of
large $\tan \beta$ and large $\sin (\beta - \alpha)$. However, the
large $\tan \beta$ region is excluded by the perturbative
unitarity constraints and therefore we will not consider this
scenario.
\begin{figure}[t!]
  \begin{center}
    \epsfig{file=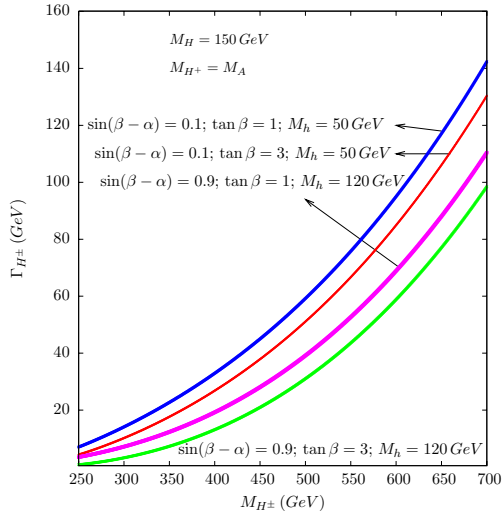,width=7 cm}
    \caption{Total width of the charged Higgs state $H^{\pm}$ for the parameter values presented
    in Figs.~\ref{plotHpmdecays1} and \ref{plotHpmdecays2}.}
    \label{plotHpmwidth}
  \end{center}
\end{figure}
In Fig.~\ref{plotHpmwidth} we show the total $H^{\pm}$ width for
the situations presented in Figs.~\ref{plotHpmdecays1} and
\ref{plotHpmdecays2}. It is clear from the plot that the width
does not depend so much on the parameters as happened for the
previous cases. This is due to the fact that all channel types are
already allowed starting from $M_{H^{\pm}} = 250 \, GeV$.

\noindent For the mass regions considered in the plots, all MSSM
scenarios predict an almost 100 \% decay to $t b$. As stated in
the introduction, the $H^\pm\to W^\pm h$ channel although
kinematically possible in the MSSM, only occurs with sizable rates
in a $\tan\beta$ region which is already excluded by experimental
data. Therefore to distinguish a charged Higgs from 2HDM we should
look for sizeable final states with one $W$ boson and some other
scalar.

\section{Conclusions}
\label{sec:conclusions} \noindent We have highlighted that in a
Type II 2HDM there exist Higgs-to-Higgs decays which are prevented
from occurring in its SUSY version, the MSSM, owing to the fact
that the latter imposes stringent relations amongst the masses of
the $H$, $A$ and $H^\pm$ states, so that they are degenerate in
mass over most of the parameter space. As these modes typically
involve decaying Higgs states that are rather heavy, they could be
primary means available at the LHC (and much less so at the
Tevatron) to dispell the MSSM hypothesis that assumes that the
sparticle states are  very heavy and beyond the kinematical reach
of the collider. An analysis of Higgs pair production in the same
spirit is now also in progress \cite{progress}.

\appendix

\section{Yukawa couplings of a Type II 2HDM}
\noindent
In this Appendix we present the Feynman rules for the Type II 2HDM Yukawa
couplings. Hereafter, the label $u$ refers to up-type quarks and neutrinos whilst $d$ to
down-type quarks and leptons. Also notice that the Goldstone bosons couple just like in
the SM, so we do not report their fermionic interactions here. Finally,
we define $\gamma_L = (1- \gamma_5)/2$ and
$\gamma_R=(1+\gamma_5)/2$. Using notation already introduced (apart from $V_{ij}$ being
the Cabibbo-Kobayashi-Maskawa matrix element in the quark sector and equating to 1 in the
lepton case), one has:
\begin{tabular}{lclclll}
\\$\overline{u}_i u_i h$: & & $-\frac{ig}{2 M_W}\frac{\cos
\alpha}{\sin \beta} \, m_{u_i}$ &
$\qquad \qquad\overline{d}_i d_i h$: & & $\frac{ig}{2 M_W}\frac{\sin \alpha}{\cos \beta} \, m_{d_i}$ \\ \\
$\overline{u}_i u_i H$: & & $-\frac{ig}{2 M_W}\frac{\sin
\alpha}{\sin \beta} \, m_{u_i}$ &
$ \qquad \qquad\overline{d}_i d_i H$: & & $-\frac{ig}{2 M_W}\frac{\cos \alpha}{\cos \beta} \, m_{d_i}$ \\ \\
$\overline{u}_i u_i A$: & & $-\frac{g}{2 M_W} \cot \beta \, m_{u_i}
\gamma_5$ & $\qquad \qquad\overline{d}_i d_i A$: & & $-\frac{g}{2
M_W} \tan \beta \, m_{d_i}
\gamma_5$ \\ \\
$\overline{u}_i d_j H^+$: & & $\frac{ig}{\sqrt{2}M_W} V_{ij} \left[
\tan \beta \, m_{d_j} \gamma_R + \cot \beta \, m_{u_i} \gamma_L
\right]$ & $ \qquad \qquad\overline{d}_i u_j H^-$: & &
$\frac{ig}{\sqrt{2}M_W} V_{ij}^* \left[ \tan \beta \, m_{d_i}
\gamma_L + \cot \beta \, m_{u_j} \gamma_R \right]$
\\\\
\end{tabular}

\noindent {\bf Acknowledgments}~ The authors thank Koji Tsumura
and Mayumi Aoki for useful discussions. Also, they acknowledge
financial support from the PMI2 Connect Initiative in the form of
a Research Co-operation Grant. SM thanks SK, YM and KY for
hospitality in Toyama while the latter thank the former for the
same reason in Southampton. RS is supported by the FP7 via a Marie
Curie Intra European Fellowship, contract number
PIEF-GA-2008-221707.

\end{document}